\documentclass[conference]{IEEEtran}
\IEEEoverridecommandlockouts
\usepackage{cite}
\usepackage{amsmath,amssymb,amsfonts}
\usepackage{algorithm}
\usepackage{algorithmic}

\usepackage{graphicx}
\usepackage{subcaption} % For subfigures
\usepackage{textcomp}
\usepackage{xcolor}
\usepackage{booktabs}
\usepackage{multirow}
\usepackage{array}
\usepackage{float}
\usepackage{manyfoot}
\usepackage{comment}
\usepackage{hyperref}
\usepackage{xcolor}
\usepackage{capt-of}
\usepackage{algorithmic}
\usepackage{svg}
\usepackage{titlesec}
\usepackage{caption}
\usepackage{tikz}
\usepackage[skip=3pt]{caption}  % Adjusts space between figure and caption

\titlespacing*{\section}{0pt}{3pt}{3pt} % {left}{before}{after}
\titlespacing*{\subsection}{0pt}{3pt}{3pt}
\titlespacing*{\subsubsection}{0pt}{3pt}{3pt}

\def\BibTeX{{\rm B\kern-.05em{\sc i\kern-.025em b}\kern-.08em
    T\kern-.1667em\lower.7ex\hbox{E}\kern-.125emX}}
\begin{document}

%Dalila, I'm suggesting to add unsupervised here? or/and maybe Prior-knowledge-free % I think unsupervised is good
\title{Hybrid Graph Embeddings and Louvain Algorithm for Unsupervised Community Detection\\
{\footnotesize \textsuperscript{}}
\thanks{}%Identify applicable funding agency here. If none, delete this.}
}
\author{\IEEEauthorblockN{Dalila Khettaf}
\IEEEauthorblockA{\textit{Computer Science Research Centre (CSRC)} \\
\textit{University of the West of England}\\
Bristol, England \\
Dalila.Khettaf@uwe.ac.uk}
\and
\IEEEauthorblockN{Djamel Djenouri}
\IEEEauthorblockA{\textit{Computer Science Research Centre (CSRC)} \\
\textit{University of the West of England}\\
Bristol, England \\
Djamel.Djenouri@uwe.ac.uk}
\and
\IEEEauthorblockN{\hspace{30mm} Zeinab Rezaeifar}
\IEEEauthorblockA{\hspace{30mm} \textit{Computer Science Research Centre (CSRC)} \\
\textit{\hspace{30mm} University of the West of England}\\
\hspace{30mm} Bristol, England \\
\hspace{30mm} zeinab.rezaeifar@uwe.ac.uk}
\and
\IEEEauthorblockN{\hspace{25mm} Youcef Djenouri}
\IEEEauthorblockA{\textit{\hspace{25mm} University of South-Eastern Norway}\\
\textit{\hspace{25mm} Norwegian Research Center}\\
\hspace{25mm} Oslo, Norway \\
\hspace{25mm} youcef.djenouri@usn.no}

}
\begin{comment}
      \author{\IEEEauthorblockN{Dalila Khettaf}
\IEEEauthorblockA{\textit{University of the West of England}\\
Bristol, UK \\
Dalila.Khettaf@uwe.ac.uk}
\and
\IEEEauthorblockN{Djamel Djenouri}
\IEEEauthorblockA{\textit{University of the West of England}\\
Bristol, UK \\
Djamel.Djenouri@uwe.ac.uk}
\and
\IEEEauthorblockN{Zeinab Rezaeifar}
\IEEEauthorblockA{\textit{University of the West of England}\\
Bristol, UK \\
zeinab.rezaeifar@uwe.ac.uk}
\and
\IEEEauthorblockN{Youcef Djenouri}
\IEEEauthorblockA{\textit{University of South-Eastern Norway}\\
\textit{Norwegian Research Center}\\
Oslo, Norway \\
youcef.djenouri@usn.no}

} 
\end{comment}

\IEEEaftertitletext{\vspace{-10mm}}
\maketitle

\begin{abstract}
% Intro sentence
% Purpose of the paper: why community detection is important
% The paper's contributions
% Paper's results compared to SOA.
% Conclusion.

This paper proposes a novel community detection method that integrates the Louvain algorithm with Graph Neural Networks (GNNs), enabling the discovery of communities without prior knowledge. Compared to most existing solutions, the proposed method does not require prior knowledge of the number of communities. It enhances the Louvain algorithm using node embeddings generated by a GNN to capture richer structural and feature information. Furthermore, it introduces a merging algorithm to refine the results of the enhanced Louvain algorithm, reducing the number of detected communities. To the best of our knowledge, this work is the first one that improves the Louvain algorithm using GNNs for community detection. The improvement of the proposed method was empirically confirmed through an evaluation on real-world datasets. The results demonstrate its ability to dynamically adjust the number of detected communities and increase the detection accuracy in comparison with the benchmark solutions. 
The code is available at \url{https://github.com/Dalila206/HGLouvain}. 
\end{abstract}


\section{Introduction}
Community detection is fundamental for graph analysis and has many real-world applications. It enables the identification of specific groups that represent meaningful contexts within a network. For example, in web applications, grouping clients based on similar interests or geographic proximity can enhance service delivery \cite{fortunato2010community}. As a general abstraction, community detection enables the clustering of similar vertices and uncovering the hierarchical structure of a graph. 
Traditional community detection methods, such as the Louvain algorithm \cite{Blondel2008Fast}, %Dalila, put the original ref of Louvain algo here %done
rely solely on the graph topology and overlook valuable node features that can enhance community identification. While the graph topology provides good insights into how vertices are connected in the graph, it does not provide any information about the node attributes. %Moreover, real-world graphs are often irregular, as vertices vary significantly in their degree of connectivity.

More recent community detection algorithms leverage both graph topology and node attributes for better vertices' separation \cite{shchur2019overlapping} \cite{Yuan2022Community} \cite{chen2017supervised}. However, because of the size of node attributes, they are usually compacted into a low-dimensional form known as node embeddings. Numerous methods have been proposed to calculate node embeddings (also known as the latent representation of vertices), such as DeepWalk \cite{Perozzi2014DeepWalk}, node2vec \cite{Leskovec2016node2vec}, and TransE \cite{Bordes2013Translating}. However, these shallow methods do not generalize well to unseen data. The more recent algorithms are based on deep learning (DL) \cite{Srichandra2024Community}. An ideal node embedding preserves the proximity of nodes, i.e., nodes that are closely connected in the graph structure have embeddings that are similarly close in the feature space. This property is well-aligned with community detection, where the spatial relationships between nodes in the feature space reflect meaningful groupings within the graph \cite{rozemberczki2019gemsec}. Node embeddings capture node-level representations but might fail to fully exploit graph structure. 

To leverage both graph topology and node features, we propose an unsupervised method GNLouv, that improves the Louvain algorithm \cite{Blondel2008Fast} by incorporating the embeddings of nodes generated by a graph neural network(GNN) in the community detection process. GNLouv combines the Louvain algorithm, which identifies the structure of the graph, with GNNs, which learn detailed representations of the nodes. These node representations are incorporated into the objective function of the Louvain algorithm to improve its results. To the best of our knowledge, we present the first paper that tackles community detection using GNNs and the Louvain algorithm. The main contributions of this article are listed as follows: 

\begin{itemize}
    \item We integrate the Louvain algorithm with node embeddings, allowing the number of communities to be determined dynamically without prior knowledge. To further refine the results, we introduce a merging algorithm based on the Louvain method, designed to reduce the number of detected communities.
    
    \item We propose a novel objective function for the Louvain algorithm that integrates both modularity and node embeddings. This allows the Louvain algorithm to detect more robust and meaningful communities by considering both graph structure and node features.  

    \item GNLouv undergoes rigorous evaluation against well-known community detection approaches, demonstrating superior capability in capturing the relevant features needed for the community detection process.
\end{itemize}

%The rest of this article is organized as follows: Section \ref{sec:lit} presents the literature review on community detection using the Louvain algorithm and GNNs. Section \ref{sec:cont} details the proposed solution as well as the datasets, the metrics used, and the results. Section \ref{sec:disc} discusses the results, and finally section \ref{sec:conc} concludes the paper.
% Contributions and soa

% Results and paper sections.
% Dalila, a literature review might be appropriate for description or long report title. In papers we generally use Related work

\section{Related work} \label{sec:lit}
Community detection has evolved throughout the years, starting with methods based on graph partitioning in the 70s to DL techniques now. Early methods relied on using the graph topology to detect communities. They include different classes of algorithms, such as graph partitioning \cite{Kernighan1970An} \cite{Fiduccia1988An} \cite{Karypis1999Multilevel}, hierarchical clustering \cite{Sibson1973SLINK} \cite{Rohlf1973Algorithm} \cite{Defays1977An}, modularity-based algorithms \cite{Newman2004Fast} \cite{Blondel2008Fast} \cite{Traag2019From}, and random walks \cite{Ballal2022Network} \cite{Hughes1996Random} \cite{Zhou2004Network}, \cite{fortunato2010community}. The rest of this section includes a review of 1) methods that improve the Louvain algorithm and 2) those using node embeddings with GNNs for community detection.

\subsection{Improvements of the Louvain algorithm}

The Louvain algorithm has several issues, including an excessive number of communities, long runtimes on large networks, non-determinism (where different runs yield different community structures), and difficulty detecting smaller communities, often leading to suboptimal partitions.

Since the Louvain algorithm was proposed, many researchers have attempted to improve it. Existing trends range from traditional to DL methods. One of the main purposes of the improvement is to reduce the time complexity of the algorithm and improve modularity. For instance, Do et al. \cite{Do2024improvement} proposed a method that adds a random walk as an additional phase. This led to an increase in modularity and Normalized Mutual Information (NMI). Van et al. \cite{Van2021Hybrid} suggested using a knowledge graph to cluster content according to users' behavior. This improved the precision and the recall. Traag et al. \cite{Traag2015Faster} improved the theoretical runtime complexity from $\mathcal{O}(m)$ to $ \mathcal{O}(n log(k))$ where n is the number of nodes, m is the number of edges and k is the community size, by moving nodes to a random neighboring community, instead of the best neighboring community. %Dalila, the improvement is not clear, what's n and m? done 
Ozaki et al. \cite{Ozaki2016simple} proposed Louvain Prune that reduces the computational time by 90\% due to the improved calculation of the change in modularity. Hu et al. \cite{Hu2016Improving} improved the runtime by combining the Louvain algorithm with the LPA algorithm when processing massive data.  Checconi et al. \cite{Checconi2015Scalable} scaled the Louvain algorithm to large networks by using a different graph mapping and data representation. Similarly, Wickramaarachchi et al. \cite{Wickramaarachchi2014Fast} proposed a distributed memory parallel algorithm that parallelizes the first few iterations, which are the most costly. Their method is five times faster than the basic Louvain algorithm. Traag et al. \cite{Traag2019From} fixed one of the main issues of the Louvain algorithm to detect more connected communities. 
These works primarily integrate traditional algorithms like LPA and graph partitioning methods into the Louvain algorithm, aiming to enhance modularity and improve its overall runtime. However, most do not focus on accuracy.

\subsection{GNNs for community detection}
Sobolevsky et al. \cite{sobolevsky2022graph} proposed a framework that combines GNNs and modularity optimization for unsupervised community detection by using a recurrent neural network to optimize modularity as part of the activation function. The method was evaluated against techniques such as Louvain \cite{Blondel2008Fast}, Leiden \cite{Traag2019From}, and Combo \cite{Sobolevsky2014General}, which showed superior or Benchmark-level modularity scores.
Bruna et al. \cite{Bruna2017community} studied data-driven methods for community detection in graphs. They studied a GNN in the synthetic Stochastic Block Model to prove that the model can reach statistical and computational detection thresholds. The GNN was demonstrated to be efficient by comparing it against the Community-Affiliation Graph Model (AGM) where it showed superior accuracy on SNAP datasets. %% accuracy only --> inspire from
Srichandra et al. \cite{Srichandra2024Community} proposed a method that combines a graph attention network(GAT) with two clustering algorithms, Kmeans and agglomeration. The GAT generates embeddings of the network, while the clustering algorithm uses the knowledge to cluster the data. The method was compared to Louvain and Girvan Newman algorithms, where it showed superior performance on the Cora and Citeseer datasets. % drawback
% pretty close to your work, non overlapping communities
Yuan et al. \cite{Yuan2022Community} proposed to use Markov Stability as a loss function over the community affiliation weight matrix in the GNN. The technique was tested and evaluated on four network datasets and surpassed other existing methods. % no ACC, NMI and omega score
Shchur et al. \cite{shchur2019overlapping} proposed Neural Overlapping Community Detection (NOCD) that leverages a graph convolutional network(GCN) to generate the affiliation matrix \( F \) for the Bernoulli Poisson model. NOCD demonstrated superior performance in terms of Overlapping NMI and exhibited promising scalability.
Chen et al. \cite{chen2017supervised} introduced the Line GNN (LGNN), a modified GNN architecture that incorporates a non-backtracking operator on the graph. The model and its variations' performance were tested on the SNAP datasets, where they achieved the best performance.
Most of these works are based on the assumption that the number of communities is known in advance, but this is not always true in the real world. GNLouv, however, is completely unsupervised and does not make any assumption on the number of communities. 

% Ask if you need to add a table; no space
% Ask if you need drawbacks of methods; criticize at the end

\begin{comment}
  \begin{table}[htbp]
\caption{Table Type Styles}
\begin{center}
\begin{tabular}{|c|c|c|c|}
\hline
\textbf{Paper}&\multicolumn{3}{|c|}{\textbf{Metrics}} \\
\cline{2-4} 
\textbf{ } & \textbf{\textit{Table column subhead}}& \textbf{\textit{Subhead}}& \textbf{\textit{Subhead}} \\
\hline
copy& More table copy$^{\mathrm{a}}$& &  \\
\hline
\multicolumn{4}{l}{$^{\mathrm{a}}$Sample of a Table footnote.}
\end{tabular}
\label{tab1}
\end{center}
\end{table}  
\end{comment}

\section{Proposed Solution} \label{sec:cont}

%Dalila, the quality of this figure, and all the figures in the expiriment section is very bad. Try to use some profesional tools togenerate. Use larger/better font, etc.

%\textcolor{red}{Dalial: Please can you draw a sketch of the designed solution here? } done

%Dalila, I already told you avoid too much direct style, we, we we, our. I fixed here but check the rest of the paper. I think this section is about the proposed solution, not the expiriment as you said.

The proposed solution is described in this section. The term “communities” refers to communities in the graph that are aimed to be detected, while the term “classes” refers to the ground truth of the datasets. 

\begin{figure}[ht!]
\begin{tabular}{c}
\includegraphics[width=0.4\textwidth]{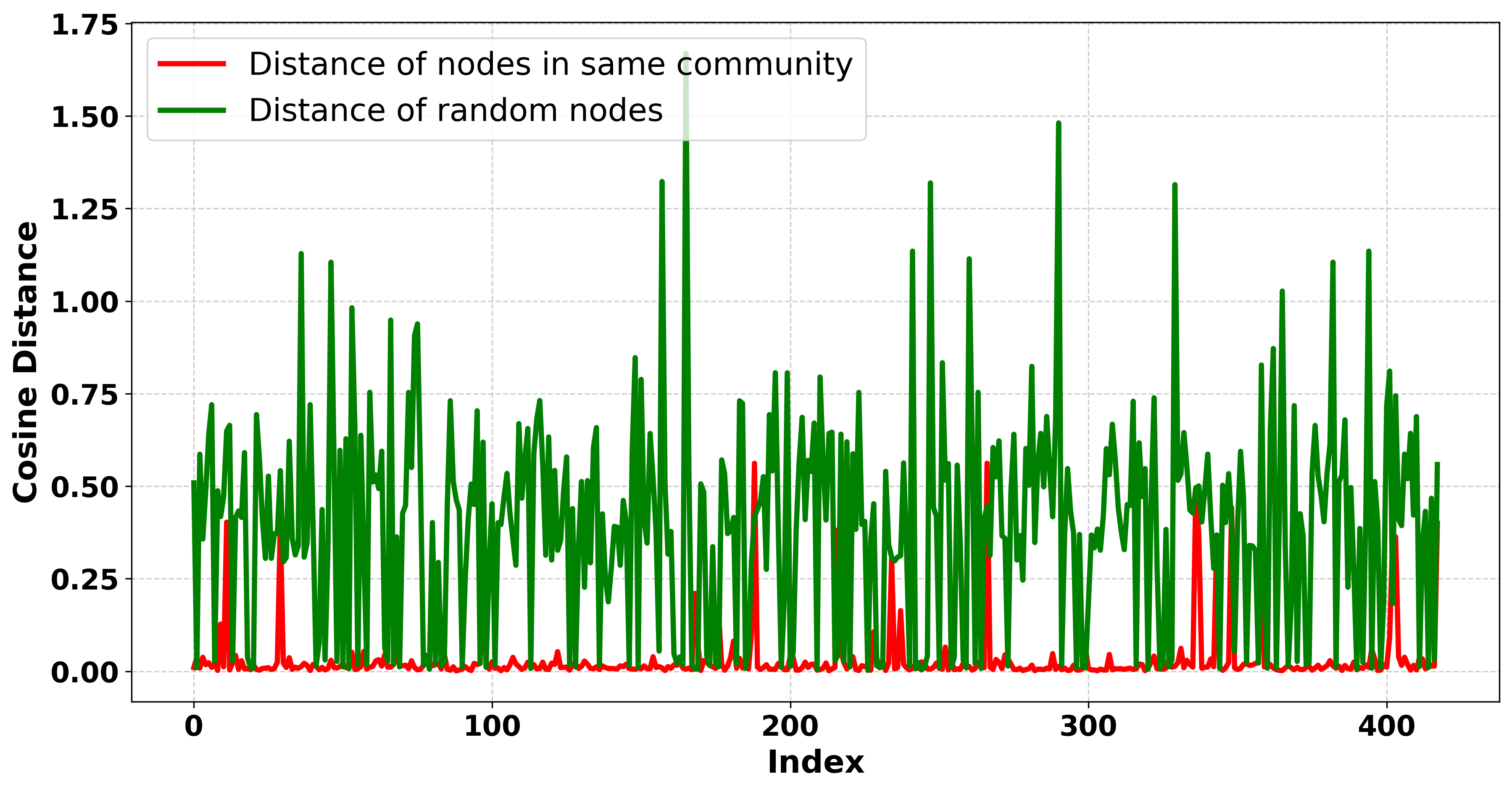} \\ 
\textit{Results on Class 2 of the Cora dataset.}\\
\includegraphics[width=0.4\textwidth]{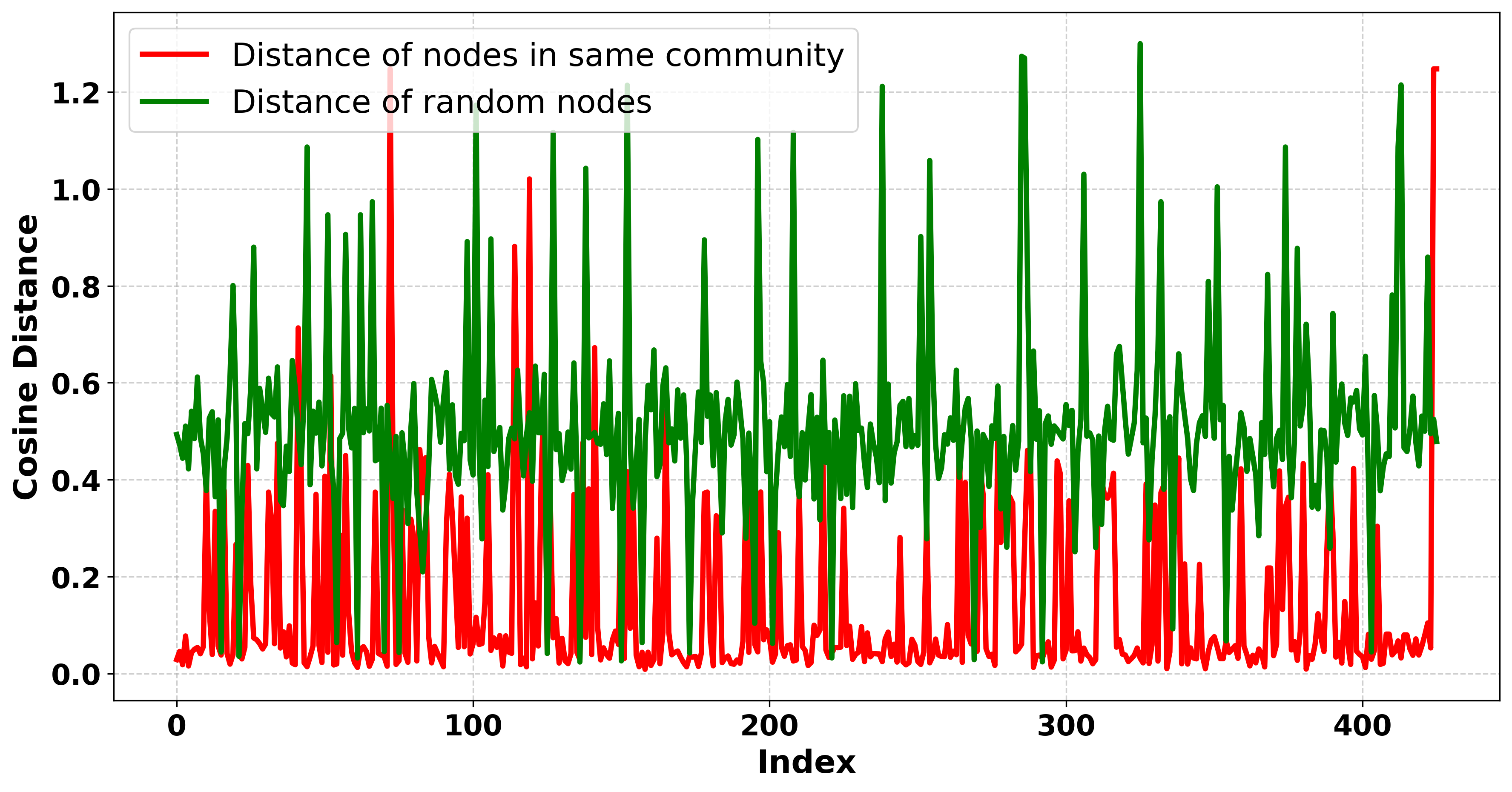}   \\
\textit{Results on Class 4 of the Cora dataset.} 
\end{tabular}
 \caption{Results on different classes of the Cora Dataset.}
\label{fig:class}
\end{figure}

\subsection{Hypothesis}
GNLouv is based on the hypothesis that vertices that belong to the same community are characterized by embeddings that are close to one another in terms of a distance metric. That is, if the vertices $ v_{1}$ and $ v_{2}$ belong to the same community, their distance $ D( v_{1},  v_{2})$ is very small. As a first step to prove that this hypothesis holds, we run some experiments. A multi-layer GCN is used to generate node embeddings, and the cosine similarity is used to calculate the distance. This test aims to prove that the distance between a community and nodes belonging to it, is smaller than the distance between the same community and nodes belonging to a different community. For example, the “book” community consists of nodes that belong to the “book” class. The distance between a random “book” and the “book” community is supposed to be smaller than the distance between a “song” and the “book” community.
To run the test, communities are created from the Cora dataset. These communities are created using $50$ random nodes from each class, that is $7$ communities in total. The distance between random nodes from the same class and their community was calculated first, and then the distance between random nodes from different communities and the community in question. A simple cosine distance formula between the embedding of the node and the centroid of the embeddings of the community in question was used for distance calculation. 
Figure \ref{fig:class} shows the test result on classes 2 and 4 of the Cora dataset. The distance between nodes in the same class appears to be very small compared to the distance from random nodes.  

\subsection{Methodology}

\begin{figure}[ht!]
    \centering    \includegraphics[width=0.5\textwidth]{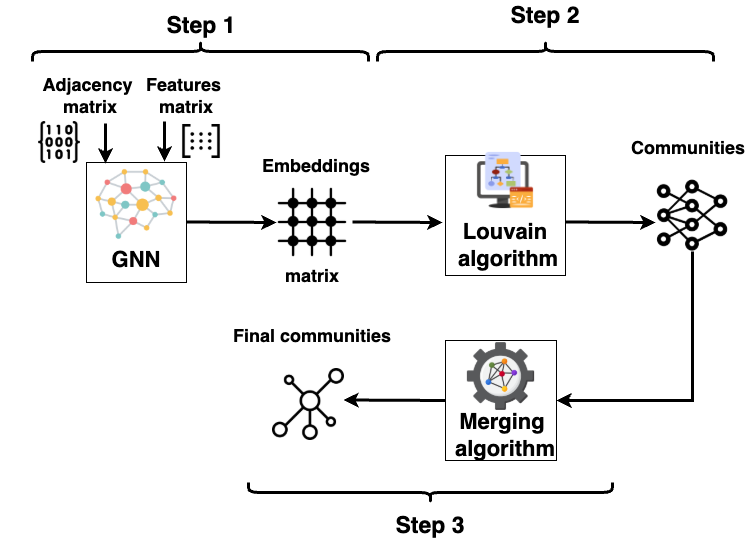} 
    \caption{Illustration of the Designed Method.}
    \label{fig:method}
\end{figure}

\begin{algorithm}[ht!]
\scriptsize
\caption{Community detection with GNLouv}
\label{algo:method}
\begin{algorithmic}[1]
\STATE \textbf{Input:} $V=\{v_1, v_2, \dots, v_m\}$: Set of nodes. $X$: The features matrix $X$, and $A$ : The adjacency matrix. 
\STATE \textbf{Output:} $C=\{C_1, C_2, \dots, C_n\}$: The set of communities.

\STATE \textbf{Step 1: Node Embeddings}
\STATE $\mathbf{E} \leftarrow GCN(X, A)$

\STATE \textbf{Step 2: Improved Louvain Algorithm}
\FOR{each node $v_i \in V$}
\STATE $C_i \leftarrow v_i$
\ENDFOR

\WHILE{modularity improved}
    \FOR{each node $v_i$}
        \STATE Calculate $\Delta_Q$
        \STATE $N_{v_i} \leftarrow NeighborCommunity(v_i)$
        \STATE $D \leftarrow Distance(\mathbf{E}_{v_i}, \overline{\mathbf{E}_{N_{v_i}}})$
        \STATE Compute $\textit{f}(Q, D)$ using Eq. \ref{eq:finalcombo}
        \STATE Move node to community with the highest $\textit{f}(Q, D)$.
    \ENDFOR
    \STATE Recalculate the graph by collapsing communities into supernodes.

\ENDWHILE

\STATE \textbf{Step 3: Iterative Community Merging}
\STATE Initialize: $T_{\text{initial}}$, $\alpha$, $T_{\text{min}}$, $It_{\text{max}}$
\WHILE{$iterations < It_{\text{max}}$}
    \FOR{each $(C_i, C_j)$}
        \IF{$D(nodes(C_i), \overline{\mathbf{E}_{C_j}}) < T_{\text{initial}}$}
            \STATE Merge($C_i$, $C_j$)
        \ENDIF
    \ENDFOR
    \IF{No communities merged}
        \STATE $T_{\text{initial}} \leftarrow T_{\text{initial}} - \alpha$  
        \IF{$T_{\text{initial}} < T_{\text{min}}$}
            \STATE Break
        \ENDIF
    \ENDIF
    \STATE $iterations \leftarrow iterations +1$
\ENDWHILE

\WHILE{Desired number of communities not reached}
    \STATE $\mathbf{E} \leftarrow Update(\mathbf{E}, C)$ 
    \STATE Repeat Step 3 
\ENDWHILE
\RETURN $C$
\end{algorithmic}
\end{algorithm}

GNLouv integrates GNNs with the Louvain algorithm to enhance the community detection task. The GNN is responsible for learning and extracting meaningful node representations, capturing both structural and feature-based relationships within the network. These learned embeddings are then utilized by the Louvain algorithm to identify communities more effectively. By leveraging the rich feature representations from the GNN, the Louvain algorithm can achieve more accurate and adaptive community detection. The overall framework of GNLouv is illustrated in Figure \ref{fig:method}, and the pseudo-code is shown in Algorithm \ref{algo:method}. It is composed of the three steps described below:

\begin{enumerate}
    \item \textit{Node Embedding}: This step consists of generating the node embeddings using the multi-layer GCN. The GCN is trained with both the adjacency matrix and the features matrix to obtain the embeddings matrix $\mathbf{E}$ that represents the embeddings of the nodes.

    \item \textit{Improved Louvain Algorithm}: Let $k$ be the node degree, where $k_{i}$ is the degree of node i and $k_{j}$ the degree of node $j$. $\delta(c_i, c_j)$ is the Kronecker delta function that equals 1 if nodes $i$ and $j$ belong to the same community and 0 otherwise, and $m$ is the total number of edges. The modularity is defined in equation Eq.\eqref{eq:modularity}. The original objective function of the Louvain algorithm measures the modularity gain $\Delta_Q$, which is the change in modularity when a node is moved to a neighboring community. The improved Louvain algorithm preserves modularity gain while incorporating the distance $D$ between a node’s embedding and the centroid of its community’s embedding. 
\begin{equation}
Q = \frac{1}{2m} \sum_{i, j} \left[ A_{ij} - \frac{k_i k_j}{2m} \right] \delta(c_i, c_j)
\label{eq:modularity}
\end{equation}

 The new objective function is denoted as $f(Q,D)$. Since $\Delta_Q$ is much larger than $D$, $D$ is reversed to obtain $\frac{1}{D}$. Then, the Logarithm with 1 added before computation is applied to both $\frac{1}{D}$ and $\Delta_Q$ to compress large values of $\Delta_Q$ and emphasize proportional relationships between $\Delta_Q$ and $D$. The logarithmic also controls the outcome of $f(Q,D)$ to avoid zero and negative values. Considering $D$ can be negative, the absolute value of both $D$ and $\Delta_Q$ is used before applying the log. The resulting expression is shown in equation \ref{eq:finalcombo}.
\begin{comment}
\begin{equation}
f(Q,D) = \Delta_Q + D
\label{eq:combo1}
\end{equation}
\end{comment}

\begin{equation}
f(Q,D) = \log_p\left( \left| \Delta_Q \right| \right) + \log_p\left( \left| \frac{1}{D} \right| \right)
\label{eq:finalcombo}
\end{equation}

\item \textit{Iterative Community Merging}: The output of the Louvain algorithm is a list of communities. In this third step, similar communities are merged. The centroid of each community is calculated, as well as the distance between each node of the community and all the centroids. The average is then compared with the initial threshold, and the two communities $C_{x}$ and $C_{y}$ are merged if the $D(C_{x}, C_{Y}) < T_{\text{initial}}$. To address the issue of distances exceeding the initial threshold, an additional parameter is introduced, $\alpha$, which gradually reduces the initial threshold over time until it reaches the specified minimum threshold $T_{\text{min}}$. This ensures that the algorithm can continue merging communities even as distances increase. The merging algorithm is run recursively until the desired number of communities is reached. 
\end{enumerate}

\section{Experiments}
In this section, we assess the effectiveness of GNLouv by experimenting with different datasets. We start by running the Louvain algorithm alone, the Leiden algorithm alone, the improved Louvain algorithm, and finally, the iterative merging algorithm.
\subsection{Metrics} % 
For this version of GNLouv, we focus on two metrics: accuracy and number of communities. The intra-community accuracy $A_{\text{intra}}(C_i)$ defined in eq. \ref{eqn:aintralet} measures the proportion of nodes $v_j$ in community $C_i$ whose class label $\text{class}(v_j)$ matches the majority class $\text{class}(C_i)$ of the community, where $|C_i|$ is the total number of nodes in $C_i$.

\begin{equation}
    A_{\text{intra}}(C_i) = \frac{|\{ v_j \in C_i \mid \text{class}(v_j) = \text{class}(C_i) \}|}{|C_i|}
\label{eqn:aintralet}
\end{equation}

The final accuracy is the inter-community accuracy, which is the average of intra-community accuracy across all communities. 

\begin{equation}
    A_{\text{inter}} = \frac{1}{k} \sum_{i=1}^k A_{\text{intra}}(C_i)
    \label{eqn:ainterlet}
\end{equation}

The other metric, the number of communities, is the resulting number of communities after running the algorithm.

\subsection{Datasets}
We have experimented on 2 citation network datasets: Cora \footnote{https://graphsandnetworks.com/the-cora-dataset/} and Citeseer \footnote{http://konect.cc/networks/citeseer/}. Information about these datasets is available in table \ref{tab:dataset_stats}.

\begin{table}[ht!]
\centering
\scriptsize
\begin{tabular}{lrrrr}
\toprule
Dataset  & \#Nodes & \#Edges & \#Features & \#Classes \\
\midrule
Cora     & 2,708   & 5,429   & 1,433      & 7        \\
Citeseer & 3,312   & 4,732   & 3,703      & 6        \\
\bottomrule
\end{tabular}
\caption{Statistics of the datasets used in the study.}
\label{tab:dataset_stats}
\end{table}

\subsection{Results}
The first experiment consists of assessing the effectiveness of the improved Louvain algorithm on both the accuracy and the community size. We simply run the Louvain algorithm, the Leiden algorithm, and the improved Louvain algorithm and report the results in Table \ref{tab:performance_eval_1}.

\begin{table}[ht!]
    \centering
    \scriptsize
    \setlength{\tabcolsep}{8pt} % Adjust cell padding
    \renewcommand{\arraystretch}{1.2} % Adjust row spacing    
    \begin{tabular}{@{}lcc@{}}
        \toprule
        \textbf{Methods} & \textbf{Cora} & \textbf{Citeseer}  \\ 
        \midrule 
         Louvain & 91.69\%(102) & 85.23\%(469)  \\ 
         Leiden  & 91.78\%(106)  & 85.17\%(471)  \\ 
        Improved Louvain & 96.33\%(78) & 86.44\%(438)\%\\ 
        %Dalila, it's not formal to pur ours here. Give the proposed solution a short name (abreviation) and put her and everywhere you refer to it.%done
        \bottomrule
    \end{tabular}
    \caption{A performance evaluation based on accuracy and community size (between parentheses).}
    \label{tab:performance_eval_1}
\end{table}

For the Cora dataset, the improved Louvain algorithm shows a clear improvement in terms of accuracy and community size. The community size dropped from 102 and 106 communities for the Louvain and Leiden algorithms, respectively, to 78 communities. This also showed an increase of about 5\% in terms of accuracy. For the Citeseer dataset, the improved Louvain algorithm showed a slight increase in accuracy of about 1\% and a decrease in the community size from 469 and 471 communities for the Louvain and the Leiden algorithm, respectively, to 438 communities. We conclude that the improved Louvain algorithm shows an improvement in terms of accuracy and community size, however, the community size can be further refined to better approximate the true number of communities in the dataset. In the second experiment, the goal is to assess the effectiveness of the iterative merging algorithm. We run the algorithm using different values for the initial threshold, ranging from small, such as 0.05, to large, such as 0.9. The impact of such thresholds on the accuracy and the community size is reported in the figures below.

\begin{figure}[ht!]
    \centering
\includegraphics[width=0.4\textwidth]{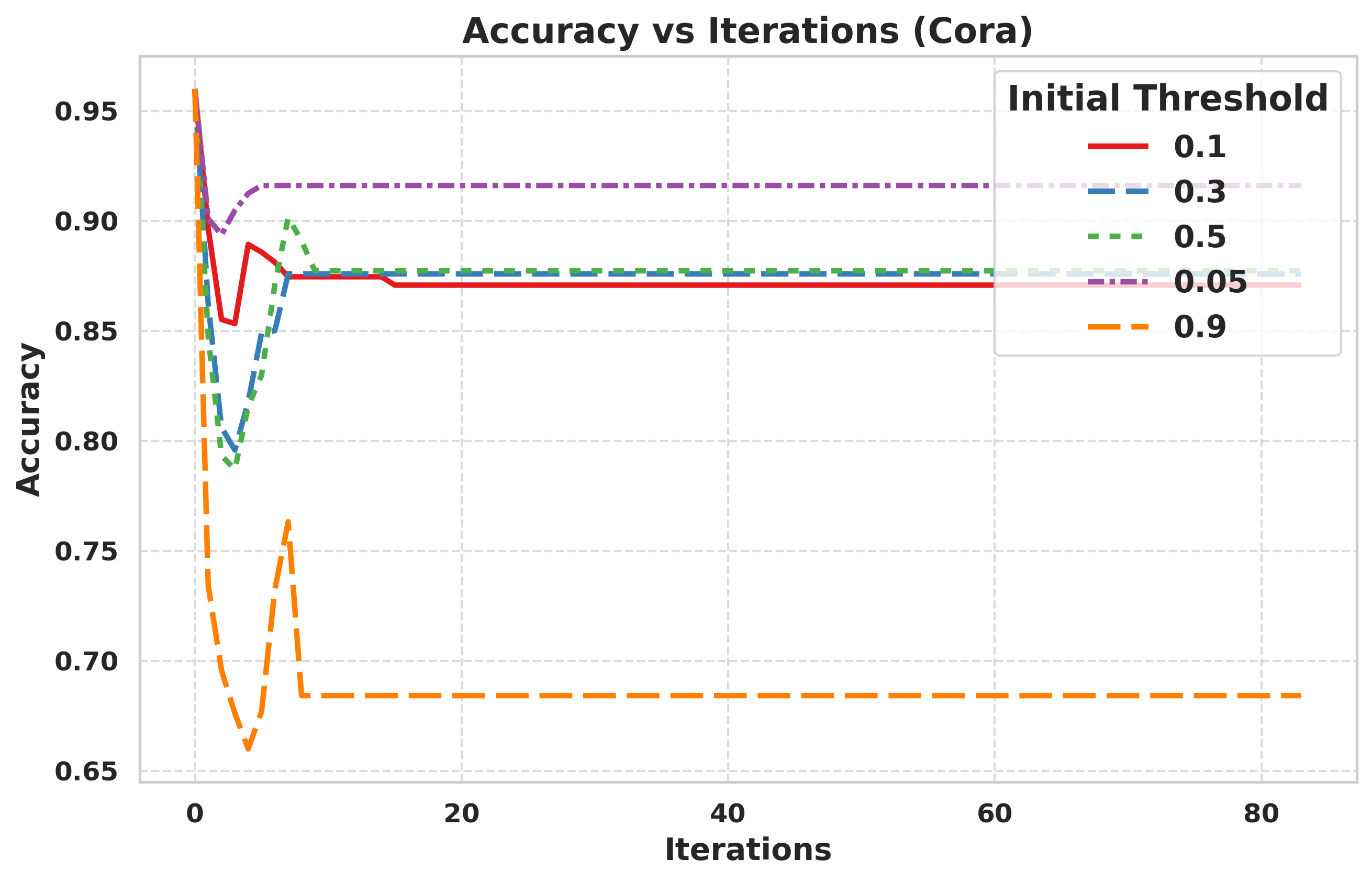}  % Adjust width as needed
    \caption{Accuracy on the Cora dataset.}
    \label{fig:corra_acc}
\end{figure}

\begin{figure}[ht!]
    \centering    \includegraphics[width=0.4\textwidth]{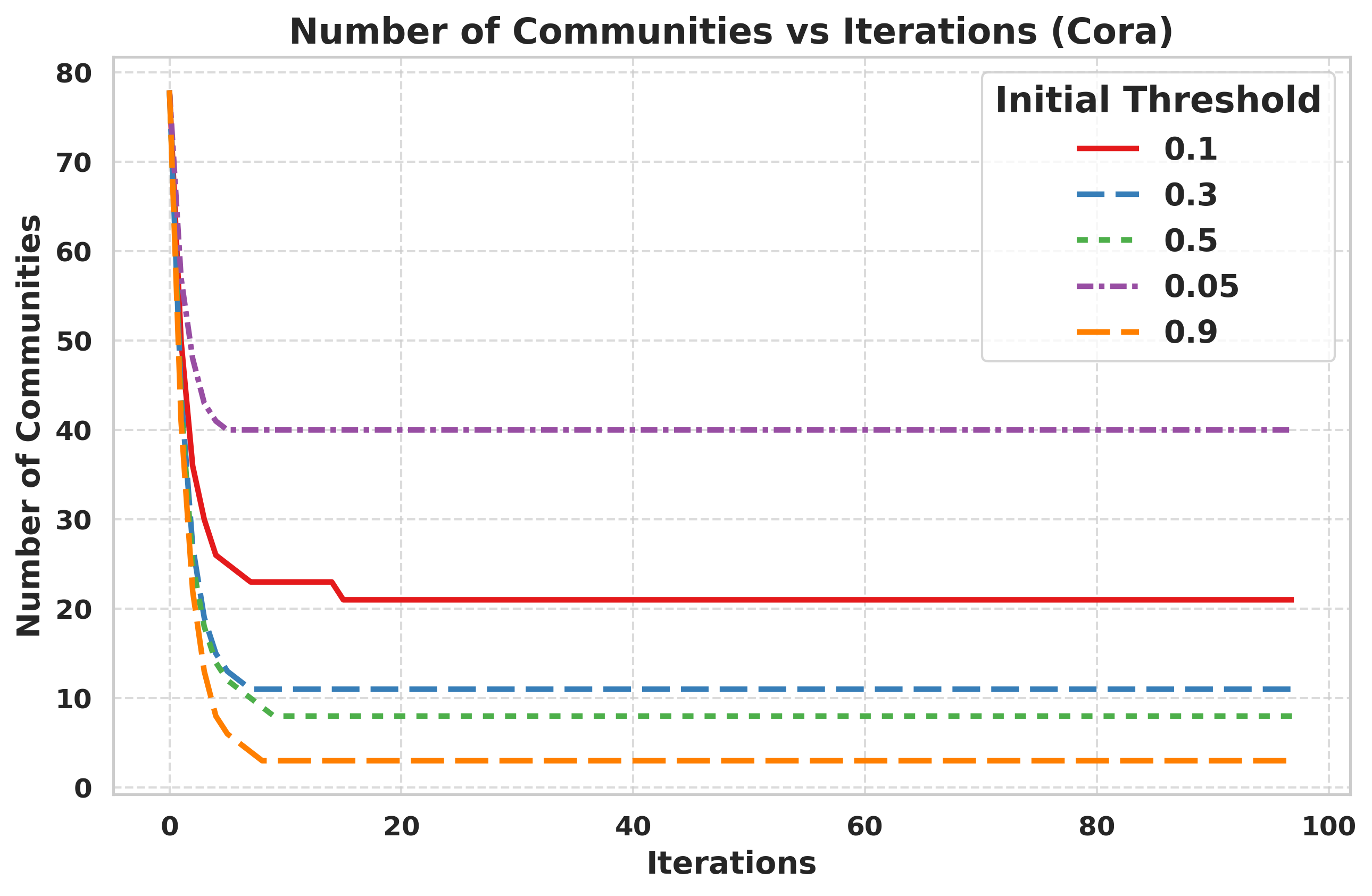}  % Adjust width as needed
    \caption{Community Size on the Cora dataset.}
    \label{fig:corra_size}
\end{figure}

\begin{figure}[ht!]
    \centering
\includegraphics[width=0.4\textwidth]{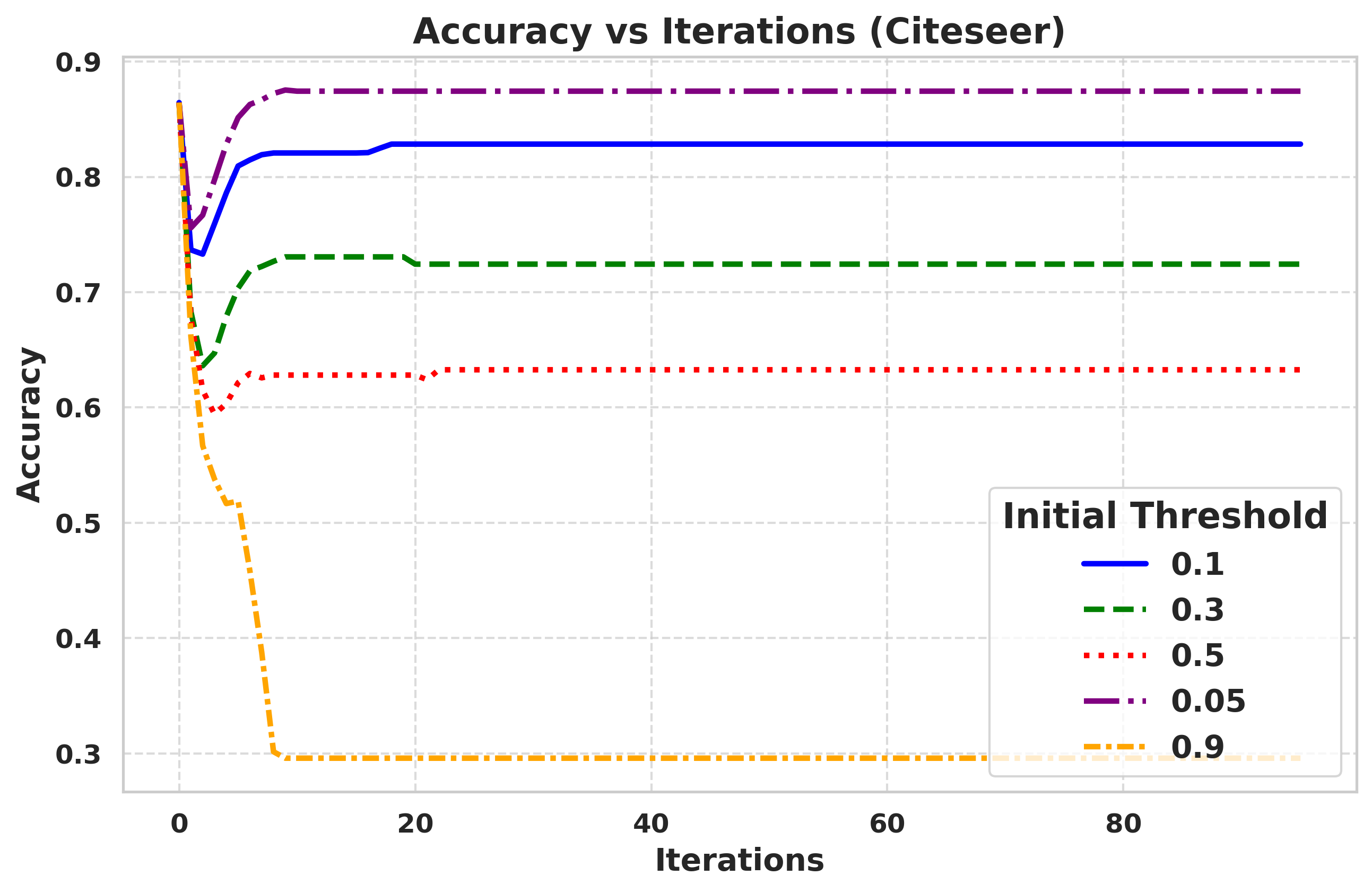}  % Adjust width as needed
    \caption{Accuracy on the Citeseer dataset.}
    \label{fig:cite_acc}
\end{figure}

\begin{figure}[ht!]
    \centering
\includegraphics[width=0.4\textwidth]{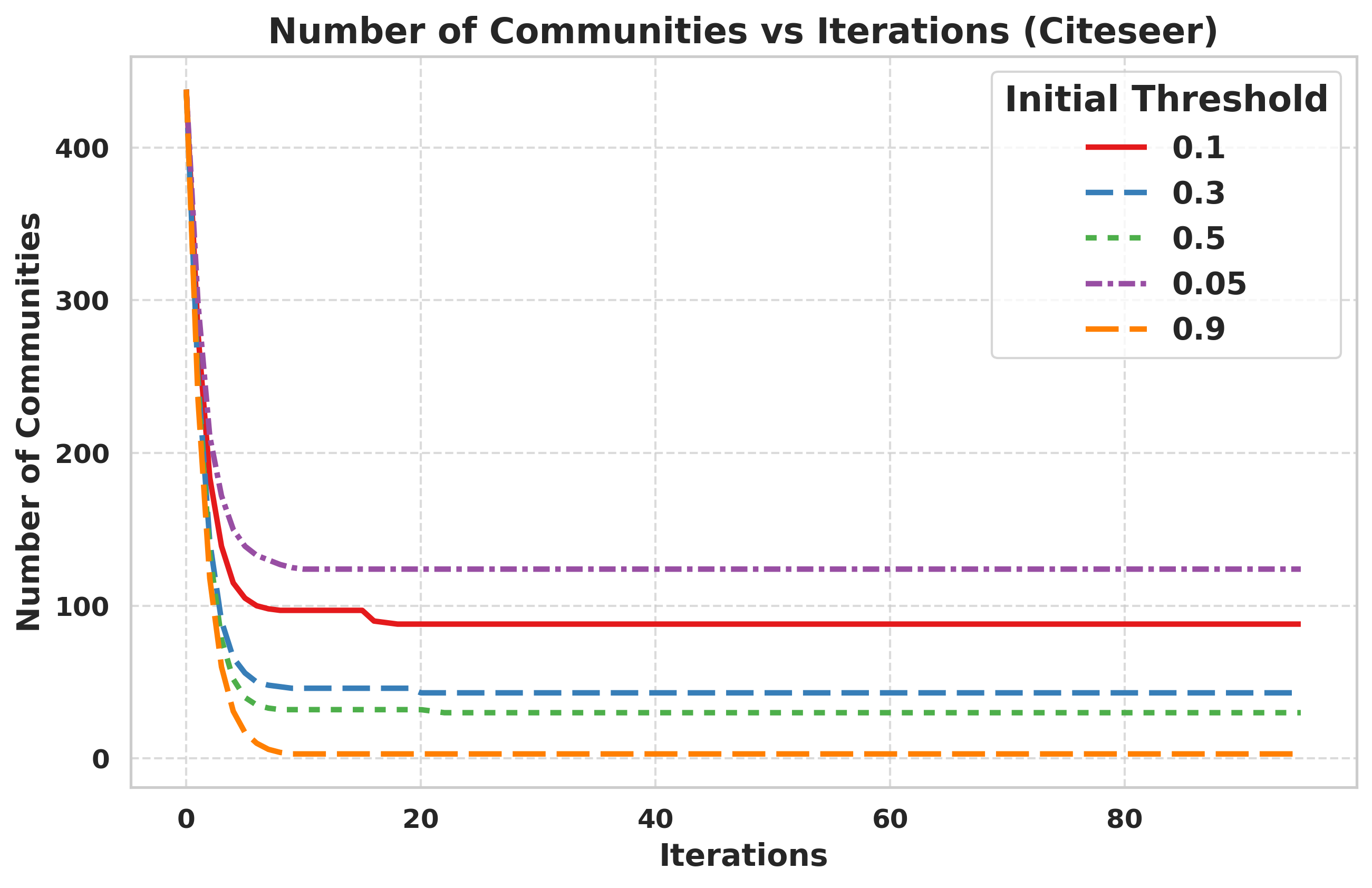}  % Adjust width as needed
    \caption{Community Size on the Citeseer dataset.}
    \label{fig:cite_size}
\end{figure}

Figures \ref{fig:corra_acc} and \ref{fig:corra_size} show both accuracy and community size for the Cora dataset. A large threshold like 0.9 results in the lowest number of communities, 3, and also results in the lowest accuracy, 68\%. The threshold 0.5 has resulted in the best community size of 8, which is close enough to the real number of communities, which is 7, and a good accuracy of 87.73\%. The lowest threshold, 0.05 has resulted in the highest accuracy 91.61\%, however, it has also resulted in the largest number of communities, 40.

Figures \ref{fig:cite_acc} and \ref{fig:cite_size} show both accuracy and community size for the Citeseer dataset. The largest threshold, 0.9, has led to the lowest accuracy 29.59\% and the lowest community size, 3. The threshold 0.05 provided the best accuracy 87.43\%, however, the community size is very large at 124. We conclude that the threshold controls the resulting accuracy and the community size. For larger thresholds, the community size is closer to the real number, however, it reduces the accuracy drastically. Smaller thresholds lead to a larger number of communities but also a higher accuracy.  

%\section{Discussion} \label{sec:disc}
%The improved Louvain algorithm has shown a slight improvement in accuracy and a reasonable decrease in the community size. While the improvements enhance the method, they are not sufficient for it to function as a standalone community detection approach.

%The communities merging algorithm provides better results based on the chosen threshold. Achieving higher accuracy often requires compromising on community size, and vice versa. While these results are preliminary, fine-tuning the parameters of the community merging algorithm suggests that an optimized version could strike a balance between accuracy and community size, preserving both effectively. 

%For more realistic scenarios, this algorithm is well-suited to dynamically control community size based on the specific problem at hand, as the optimal community size is rarely known in advance.

\section{Conclusion} \label{sec:conc}  
This work explored community detection by leveraging both node embeddings and network structure. The improved Louvain algorithm demonstrated a slight accuracy improvement and a reasonable reduction in community size, though it remains insufficient as a standalone approach. To address this, we introduced a community merging algorithm that balances accuracy and community size based on a chosen threshold. The results indicate that achieving higher accuracy often requires a trade-off with community size, but fine-tuning the merging algorithm’s parameters can optimize this balance effectively.  Moreover, the adaptability of GNLouv makes it well-suited for real-world scenarios where the optimal community size is unknown. Future work will focus on refining parameter selection to enhance performance and exploring alternative methods that assume a predefined community size to improve accuracy further. Another interesting direction is to explore the present work in real applications, including intrusion detection, and maritime surveillance for security and surveillance. Limitations of the approach include scalability on large networks, which can be explored for use on real-world networks.

\section*{Acknowledgments}
This research was mainly funded by a PhD Scholarship from the Computer Science Research Center, College of Arts, Technology and Environment, UWE Bristol. 

%\subsection{Maintaining the Integrity of the Specifications}

\begin{comment}
\begin{figure}[h]
    \centering
    \includesvg[width=0.8\textwidth]{figures/Diagramme sans nom.drawio.svg}  % Adjust width as needed
    \caption{Your SVG figure}
    \label{fig:svg_example}
\end{figure}
\end{comment}

\vspace{12pt}
\color{red}
%IEEE conference templates contain guidance text for composing and formatting conference papers. Please ensure that all template text is removed from your conference paper prior to submission to the conference. Failure to remove the template text from your paper may result in your paper not being published.


\begin{thebibliography}{00}

%%%%%%%%%%%MY papers %%%%%%%%%%%%%%%%%%%%%%%%%%%%%%%%%%%%%%%%%%%%%%%%%%%
\bibitem{sobolevsky2022graph} S. Sobolevsky and A. Belyi, “Graph neural network inspired algorithm for unsupervised network community detection,” Applied Network Science, vol. 7, no. 1, Sep. 2022.
\bibitem{Blondel2008Fast} V. D. Blondel, J.-L. Guillaume, R. Lambiotte, and E. Lefebvre, “Fast unfolding of communities in large networks,” Journal of Statistical Mechanics: Theory and Experiment, vol. 2008, no. 10, p. P10008, Oct. 2008
\bibitem{Traag2019From} V. A. Traag, L. Waltman, and N. J. van Eck, “From Louvain to Leiden: guaranteeing well-connected communities,” Scientific Reports, vol. 9, no. 1, Mar. 2019
\bibitem{Sobolevsky2014General}S. Sobolevsky, R. Campari, A. Belyi, and C. Ratti, “General optimization technique for high-quality community detection in complex networks,” Physical Review E, vol. 90, no. 1, Jul. 2014
\bibitem{Bruna2017community}J. Bruna and X. Li, “Community Detection with Graph Neural Networks,” arXiv (Cornell University), May 2017
\bibitem{Srichandra2024Community}[1] I. V. Srichandra and P. Bhadra, “Community detection using graph attention networks clustering algorithm,” 2024 IEEE 9th International Conference for Convergence in Technology (I2CT), vol. 246, pp. 1–6, Apr. 2024. 
\bibitem{Yuan2022Community} S. Yuan, C. Wang, Q. Jiang, and J. Ma, “Community detection with graph neural network using Markov stability,” 2022 International Conference on Artificial Intelligence in Information and Communication (ICAIIC), vol. 103, pp. 437–442, Feb. 2022. 
\bibitem{shchur2019overlapping} O. Shchur and S. Günnemann, "Overlapping community detection with graph neural networks," *arXiv preprint arXiv:1909.12201*, 2019.
\bibitem{chen2017supervised} Z. Chen, X. Li, and J. Bruna, "Supervised community detection with line graph neural networks," arXiv preprint arXiv:1705.08415, 2017.
\bibitem{Fortunato2007community}Fortunato, S. and Castellano, C. 2007. Community structure in graphs. arXiv preprint arXiv:0712.2716.
\bibitem{Fortunato2010community}Fortunato, S. 2010. Community detection in graphs. Physics Reports. 486, 3-5 (2010), 75–174.
\bibitem{li2024community}Y. Li, Y. Tang, J. Cheng, C. He, and F. Tang, "A community detection method to counter the semantic noise of complex networks: Bridging a topology and semantic subspace transformation," IEEE Syst. Man Cybern. Mag., vol. 10, no. 2, pp. 2-14, 2024.
\bibitem{rashnodi2024community}O. Rashnodi, M. Rastegarpour, P. Moradi, and A. Zamanifar, "Community detection in attributed social networks using deep learning," J. Supercomput., vol. 80, no. 18, pp. 25933-25973, 2024.
\bibitem{patil2024community}J. H. Patil, P. Potikas, W. B. Andreopoulos, and K. Potika, "Community detection using deep learning: Combining variational graph autoencoders with Leiden and K-truss techniques," Inf., vol. 15, no. 9, p. 568, 2024.
\bibitem{li2024contrastive}Y. Li, J. Chen, C. Chen, L. Yang, and Z. Zheng, "Contrastive deep nonnegative matrix factorization for community detection," in ICASSP 2024-2024 IEEE Int. Conf. Acoust., Speech Signal Process. (ICASSP), 2024, pp. 6725-6729.
\bibitem{wu2024deep}X. Wu, W. Lu, Y. Quan, Q. Miao, and P. G. Sun, "Deep dual graph attention auto-encoder for community detection," Expert Syst. Appl., vol. 238, p. 122182, 2024.
\bibitem{chai2024deep} B. Chai, Z. Li, and X. Zhao, "Deep graph clustering by integrating community structure with neighborhood information," Information Sciences, vol. 120951, 2024.
\bibitem{zhang2024feature}L. Zhang, Z. Wu, H. Yang, W. Zhang, and P. Zhou, "Feature Graph Augmented Network Representation for Community Detection," *IEEE Transactions on Computational Social Systems*, 2024.
\bibitem{shan2024graph}Z. Shan, D. Zhang, and Y. Lei, "Graph contrastive learning with cross-encoder for community discovery," *Applied Intelligence*, vol. 54, no. 2, pp. 2211–2224, 2024.
\bibitem{sun2020network}H. Sun, F. He, J. Huang, Y. Sun, Y. Li, C. Wang, and X. Jia, "Network embedding for community detection in attributed networks," *ACM Transactions on Knowledge Discovery from Data (TKDD)*, vol. 14, no. 3, pp. 1–25, 2020.
\bibitem{rozemberczki2019gemsec} B. Rozemberczki, R. Davies, R. Sarkar, and C. Sutton, "Gemsec: Graph embedding with self clustering," 2019 IEEE/ACM International Conference on Advances in Social Networks Analysis and Mining (ASONAM), Vancouver, BC, Canada, 2019, pp. 65–72.
\bibitem{newman2004finding}M. E. Newman and M. Girvan, "Finding and evaluating community structure in networks," Physical Review E, vol. 69, no. 2, p. 026113, Feb. 2004.
\bibitem{fortunato2010community}S. Fortunato, "Community detection in graphs," Physics Reports, vol. 486, no. 3-5, pp. 75-174, 2010.
\bibitem{Kernighan1970An}B. W. Kernighan and S. Lin, "An efficient heuristic procedure for partitioning graphs," *Bell System Technical Journal*, vol. 49, pp. 291–307, 1970.
\bibitem{Fiduccia1988An}C. M. Fiduccia and R. M. Mattheyses, "A linear-time heuristic for improving network partitions," in *Papers on Twenty-Five Years of Electronic Design Automation*, pp. 241–247, 1988.
\bibitem{Karypis1999Multilevel}G. Karypis and V. Kumar, "Multilevel k-way hypergraph partitioning," in *Proceedings of the 36th Annual ACM/IEEE Design Automation Conference*, June 1999, pp. 343–348.
\bibitem{Sibson1973SLINK}R. Sibson, "SLINK: an optimally efficient algorithm for the single-link cluster method," *The Computer Journal*, vol. 16, no. 1, pp. 30–34, 1973.
\bibitem{Rohlf1973Algorithm}F. Rohlf, "Algorithm 76. Hierarchical clustering using the minimum spanning tree," *The Computer Journal*, vol. 16, pp. 93–95, 1973.
\bibitem{Defays1977An}D. Defays, "An efficient algorithm for a complete link method," *The Computer Journal*, vol. 20, no. 4, pp. 364–366, 1977.
\bibitem{Newman2004Fast}M. E. Newman, "Fast algorithm for detecting community structure in networks," *Physical Review E—Statistical, Nonlinear, and Soft Matter Physics*, vol. 69, no. 6, p. 066133, 2004.

\bibitem{Ballal2022Network}A. Ballal, W. B. Kion-Crosby, and A. V. Morozov, "Network community detection and clustering with random walks," Phys. Rev. Res., vol. 4, no. 4, p. 043117, 2022.
\bibitem{Hughes1996Random}B. D. Hughes, Random Walks and Random Environments, Oxford University Press, 1996.
\bibitem{Zhou2004Network}H. Zhou and R. Lipowsky, "Network brownian motion: A new method to measure vertex-vertex proximity and to identify communities and subcommunities," Lect. Notes Comput. Sci., vol. 3038, pp. 1062–1069, 2004.
\bibitem{Perozzi2014DeepWalk}B. Perozzi, R. Al-Rfou, and S. Skiena, "DeepWalk: Online learning of social representations," in Proc. 20th ACM SIGKDD Int. Conf. Knowledge Discovery and Data Mining, Aug. 2014, pp. 701–710.
\bibitem{Leskovec2016node2vec}A. Grover and J. Leskovec, "node2vec: Scalable feature learning for networks," in Proc. 22nd ACM SIGKDD Int. Conf. Knowledge Discovery and Data Mining, Aug. 2016, pp. 855–864.
\bibitem{Bordes2013Translating}A. Bordes, N. Usunier, A. Garcia-Duran, J. Weston, and O. Yakhnenko, "Translating embeddings for modeling multi-relational data," in Advances in Neural Information Processing Systems, vol. 26, 2013.
\bibitem{Do2024improvement} D. H. Do and T. H. D. Phan, "An improvement on the Louvain algorithm using random walks," *arXiv preprint arXiv:2403.08313*, 2024.
\bibitem{Van2021Hybrid}H. Van Pham and D. N. Tien, "Hybrid Louvain-clustering model using knowledge graph for improvement of clustering user’s behavior on social networks," in *Intelligent Systems and Networks: Selected Articles from ICISN 2021, Vietnam*, Singapore: Springer, 2021, pp. 126–133.
\bibitem{Traag2015Faster}V. A. Traag, "Faster unfolding of communities: Speeding up the Louvain algorithm," *Phys. Rev. E*, vol. 92, no. 3, p. 032801, 2015.
\bibitem{Ozaki2016simple}N. Ozaki, H. Tezuka, and M. Inaba, "A simple acceleration method for the Louvain algorithm," *Int. J. Comput. Electr. Eng.*, vol. 8, no. 3, p. 207, 2016.
\bibitem{Hu2016Improving}B. Hu, W. Li, X. Huo, Y. Liang, M. Gao, and P. Pei, "Improving Louvain algorithm for community detection," in *Proc. 2016 Int. Conf. Artif. Intell. Eng. Appl.*, Nov. 2016, pp. 110–115, Atlantis Press.
\bibitem{Checconi2015Scalable}X. Que, F. Checconi, F. Petrini, and J. A. Gunnels, "Scalable community detection with the Louvain algorithm," in *Proc. 2015 IEEE Int. Parallel Distrib. Process. Symp.*, May 2015, pp. 28–37.
\bibitem{Wickramaarachchi2014Fast}C. Wickramaarachchi, M. Frincu, P. Small, and V. K. Prasanna, "Fast parallel algorithm for unfolding of communities in large graphs," in *Proc. 2014 IEEE High Perform. Extreme Comput. Conf. (HPEC)*, Sep. 2014, pp. 1–6.
\bibitem{Xiong2020Large}S. J. Xiong, S. B. Chen, C. H. Ding, and B. Luo, "Large-scale network representation learning based on improved Louvain algorithm and deep autoencoder," in Pattern Recognition and Computer Vision: Third Chinese Conference, PRCV 2020, Nanjing, China, October 16–18, 2020, Proceedings, Part III, vol. 3, pp. 446–459, Springer International Publishing, 2020.
\bibitem{Liu2024community}C. Liu, Y. Han, H. Xu, S. Yang, K. Wang, and Y. Su, "A community detection and graph-neural-network-based link prediction approach for scientific literature," Mathematics, vol. 12, no. 3, p. 369, 2024.
\bibitem{Mernyei2020Wiki}P. Mernyei and C. Cangea, "Wiki-CS: A Wikipedia-based benchmark for graph neural networks," arXiv preprint arXiv:2007.02901, 2020.






















\end{thebibliography}
\end{document}